\documentclass[conference]{IEEEtran}

\ifCLASSINFOpdf
\else
\fi
\usepackage{multicol}
\usepackage{bbm}
\usepackage{booktabs}
\usepackage{algpseudocode}
\usepackage{algorithm}

\usepackage{color}
\usepackage[cmex10]{amsmath}
\usepackage{dblfloatfix}
\usepackage{amsfonts}
\usepackage{url}
\usepackage{pgfplots}
\usepackage{pgfplotstable}
\usepackage{hyperref}
\hypersetup{
    colorlinks=true,
    linkcolor=blue,
    filecolor=magenta,      
    urlcolor=cyan,
}
\usepackage{listings}

\begin{document}
\pagenumbering{gobble}

%
\title{\textbf{\Large Interactive Process Identification and Selection from SAP ERP\\[-1.5ex]}}

\author{
\IEEEauthorblockN{~\\[-0.4ex]\large Julian Weber, Alessandro Berti\IEEEauthorrefmark{1}, Gyunam Park\IEEEauthorrefmark{1}, Majid Rafiei\IEEEauthorrefmark{1}, Wil M.P. van der Aalst\IEEEauthorrefmark{1}}
\IEEEauthorblockA{\IEEEauthorrefmark{1}Process and Data Science Department, RWTH Aachen University\\
Process and Data Science department, Lehrstuhl fur Informatik 9 52074 Aachen, Germany\\
Emails: {julian.weber1@rwth-aachen.de},{ \{a.berti, gnpark, majid.rafiei, wvdaalst \}@pads.rwth-aachen.de}}
}

\maketitle

\begin{abstract}
SAP ERP is one of the most popular information systems supporting various organizational processes, e.g., O2C and P2P.
However, the amount of processes and data contained in SAP ERP is enormous. Thus, the identification of the processes that are contained in a specific SAP instance, and the creation of a list of related tables is a significant challenge.
Eventually, one needs to extract an event log for process mining purposes from SAP ERP. This demo paper shows
the tool \emph{Interactive SAP Explorer} that tackles the process identification and selection problem by encoding the relational structure of SAP ERP in a labeled property graph.
Our approach allows asking complex process-related queries along with advanced representations of the relational structure.
\end{abstract}

\begin{IEEEkeywords}
ETL; SAP; Object-Centric Process Mining
\end{IEEEkeywords}



%
\IEEEpeerreviewmaketitle

\section{Introduction}
\label{sec:introduction}

Process mining is a branch of data science that provides methods for the analysis of event data recorded by information systems such as ERP and CRM systems.
An essential step for such analyses is extracting an event log from the information systems.
Moreover, this is one of the most time-consuming steps in most process mining projects. Thus, a successful extraction is a key to any process mining initiative.

The SAP ERP system is a popular choice for managing business processes such as order-to-cash (O2C, management of orders from the customers) and procure-to-pay (P2P, management of orders to the suppliers).
Despite its widespread adoption, the extraction of event logs from an SAP ERP system remains ad-hoc.
Commercial vendors such as Celonis provide extraction tools, focusing on the most common processes, e.g., O2C and P2P.
However, other processes are under less attention, such as inventory management,
financial planning, accounting, and production control processes, leading to few process mining projects in such processes.

It is challenging to extract event logs from SAP ERP systems due to their complexity, e.g., a typical SAP system contains 800,000 tables with tons of relationships.
This inevitably requires domain knowledge from the process experts of the organization.
To this end, the expert needs to 1) identify the process to analyze, 2) select the relevant tables containing relevant data of the process, and 3) design query statements, e.g., using SQL.

In this demo paper, we present the tool \emph{Interactive SAP Explorer} to support the domain expert for the first two initial steps, i.e., \emph{process identification} and \emph{tables selection}.
Given a user input of a core document class in the organization, the tool identifies the most relevant process and the underlying tables.
For instance, if the input by the user is \textit{purchase order document}, then the most relevant process is the P2P process, and the underlying tables are as follows: \textit{EBAN} for a purchase requisitions, \textit{EKKO} for purchase orders, \textit{EKBE} for goods/invoice receipts, \textit{RBKP/RSEG} for invoice processing, \textit{BKPF/BSEG} for payments, and \textit{CDHDR/CDPOS} for changes in documents.

The tool first encodes the relational structure of SAP in a labeled property graph inserted inside a graph database.
Then, a web interface is provided that permits the exploration of the relational structure of the SAP instance,
the identification of the most important processes,
and the creation of a list of tables for extraction. The list of tables is eventually provided to another component of the tool which has been previously introduced in \cite{DBLP:conf/icpm/BertiPRA21}
which creates an object-centric event log from such a list of tables.
The tool improves the prototype proposed in \cite{DBLP:conf/icpm/BertiPRA21} with better performance, customization, and exploration possibilities, particularly in the process identification and selection phases.

The rest of this extended abstract is organized as follows. Section \ref{sec:innovations} describes the functioning of the extractor.
Section \ref{sec:availability} points to the availability of the tool. Section \ref{sec:maturity} discusses the maturity of the tool.
Eventually, Section \ref{sec:conclusion} concludes the paper.

\section{Innovations and Features}
\label{sec:innovations}
This section explains 1) \emph{process identification and selection}, which is the main contribution of this paper, and 2) \emph{process extraction}, which uses the output of this paper to produce event logs.
First, the process identification and selection is implemented as follows:
\begin{itemize}
\item The elements of the relational structure of SAP that are important for the definition of a set of classes related to a given process are imported inside a graph database (Neo4J).
\begin{itemize}
\item A graph database permits a faster exploration of the neighboring entities to a given concept
because the connections are referenced inside the node object.
\item The chosen graph database (Neo4J) provides efficient implementations of layout algorithms, which can be executed on a significant amount of nodes/edges to provide an understandable graphical representation
of the relational structure in SAP.
\end{itemize}
\item Then, the identification process can be started. The first step is to identify a document type of interest (for example, the \emph{purchase orders} and \emph{sales orders}).
This is directly connected, in the relational structure of SAP, to a set of tables (\emph{purchase orders} are connected to the tables \emph{EKKO}, \emph{EKPO}, \emph{EKPA}, \emph{EKET}, \emph{EKKN}).
\item The next step is expanding the aforementioned set of tables. Starting from the initial set of tables, we identify the tables connected to the initial tables via the relational structure.
The union of these tables contains the set of events regarding a process in SAP. For example, by expanding the tables related to the \emph{purchase orders} document type, we get a set of tables including purchase
requisitions (\emph{EBAN}), goods/invoice receipts (\emph{EKBE}), accounting documents (\emph{BKPF}), and other tables containing the events of the P2P process in SAP.
\end{itemize}

The process extraction component, which uses the approach described in \cite{DBLP:conf/icpm/BertiPRA21}, aims to extract an object-centric event log out of the SAP system based on the relevant tables identified in the previous step. 
There is no need to specify any SQL query.

\begin{itemize}
\item A pre-processing step is performed to restrict the extraction to the desired configuration.
\item The extraction of the object-centric event log is performed, with an output following the OCEL specification \url{http://www.ocel-standard.org/}.
\end{itemize}

\section{Availability of the Application}
\label{sec:availability}

The source codes of the different components of the tool are available in the following repositories:
\begin{itemize}
\item \emph{Layer of web services that can be run on IIS}: this component can be downloaded at \url{https://github.com/Javert899/interactive-extractor-from-sap-main/tree/main/Backend-C%23/SAPExtractorAPI}.
\item \emph{Angular web application}: this component can be downloaded at \url{https://github.com/Javert899/interactive-extractor-from-sap-main/tree/main/Frontend/InteractiveSAPExtractor}.
\item \emph{Python web services for the extraction of the object-centric event log}: this component can be downloaded at \url{https://github.com/Javert899/sap-extractor}.
\end{itemize}
Note that there is a dependency on non-open source UI components which need to be licensed to a single user.
Therefore, the application is not directly runnable from the aforementioned source repositories.
Also, the extractor requires the availability of an SAP ECC instance supported by the Oracle database and the installation of the Neo4J graph database, which is released under a proprietary license.
The authors can provide access to the compiled version of the project under request.
A videocast of the application is provided at the address \url{https://www.youtube.com/watch?v=Wi2xuUS0YSY}.

\section{Maturity}
\label{sec:maturity}

The existing version of the tool can connect only to an SAP ECC instance supported by the Oracle database. Despite this being a popular option, this limits the possibility
to apply the extractor in a generic setting.
The extractor needs different components to run. This is architecturally complicated and, therefore, highly dependent on the functioning of existing queries/connectors on different
versions of the software.

Our extractor overcomes the following limitations of existing SAP extractors; they are process-specific, they rely on traditional event logs, and suffer from convergence/divergence issues.
However, there are remaining limitations, including the fairly basic definition of activity/timestamp concepts.
The choice of the graph database to navigate the relational structure of SAP is advantageous in terms of performance. After the selection of a set of tables, the extraction of an object-centric
event log is left to the Python component, which executes many SQL queries to load the information needed in memory.
Therefore, the extractor is limited by the amount of memory of the client.

The challenges are on both the theoretical and practical side. Theoretically, the selection of the activity concept is still a challenge. Practically, supporting different editions of SAP with different underlying databases, and an in-memory approach to compose the object-centric event log are still open challenges.

\section{Conclusion}
\label{sec:conclusion}

This demo paper presents an interactive extractor of object-centric event logs from SAP ERP, which is composed by two components: \emph{process identification and selection} (novelty of this paper) and \emph{process extraction} (using \cite{DBLP:conf/icpm/BertiPRA21}).
While the tool's code is open-source, it relies on some components released with a proprietary license. Section \ref{sec:maturity} discusses some limitations
of the tool with the current architecture, which compromises its applicability in an enterprise setting

.

\section{Acknowledgments}

We thank the Alexander von Humboldt (AvH) Stiftung for supporting our research. Funded by the Deutsche Forschungsgemeinschaft (DFG, German Research Foundation) under Germany's Excellence Strategy–EXC-2023 Internet of Production – 390621612.

\bibliographystyle{IEEEtran}
\bibliography{demo_icpm_sapextractor}

\end{document}